\documentclass[11pt]{article}
\usepackage{cite}
\usepackage{amsmath,amssymb,amsfonts}
\usepackage{algorithmic}
\usepackage{graphicx}
\usepackage{textcomp}
\usepackage{xcolor}
\usepackage{microtype}
\usepackage{bm}
\usepackage{amsthm}
\usepackage{url}
\usepackage{hyperref}

\newtheorem{theorem}{Theorem}

\newtheorem{corollary}{Corollary}

\newcommand{\remove}[1]{}

\usepackage{microtype}

\title{A Generalization of Teo and Sethuraman's Median Stable Marriage Theorem
\thanks{
Supported by NSF CNS-1812349, CNS-1563544, and the Cullen Trust for Higher Education Endowed Professorship}
}

\author{Vijay K. Garg,\\
  The University of Texas at Austin,\\
  Department of Electrical and Computer Engineering,\\
  Austin, TX 78712, USA}
\begin{document}
\maketitle

\begin{abstract}

Let $L$ be any finite distributive lattice and $B$ be any boolean predicate defined on $L$ such that 
the set of elements satisfying $B$ is a sublattice of $L$. Consider any subset $M$ of $L$ of size $k$ of elements
of $L$ that satisfy $B$. Then, we show that $k$ generalized median elements generated from $M$ also satisfy $B$.
We call this result generalized median theorem on finite distributive lattices.
When this result is applied to the stable matching, we get  Teo and Sethuraman's median stable matching theorem.
Our proof is much simpler than that of Teo and Sethuraman. 
When the generalized median theorem is applied to the assignment problem, we get an analogous result for market clearing price vectors.
 \end{abstract}

\pagebreak 

\section{Introduction}
The Stable Matching Problem (SMP) \cite{gale1962college} has wide applications in economics, distributed computing, resource allocation and many other fields \cite{maggs2015algorithmic,iwama2008survey} with multiple books
and survey articles\cite{gusfield1989stable,knuth1997stable,roth1992two,iwama2008survey}.
In the standard SMP, there are $n$ men, $m_1, m_2, \ldots, m_n$, and $n$ women, $w_1, w_2, \ldots, w_n$, each with their totally ordered preference list.
The goal is to find
a matching between men and women such that there is no {\em blocking pair}. A pair of a man and a woman is a blocking pair for a matching if they are not married to each other but
prefer each other over their partners in that matching.
Teo and Sethuraman \cite{teo1998geometry} have shown the following result called generalized median stable matching theorem.

\begin{theorem}\cite{teo1998geometry} 
 Let $M_1, M_2, . . ., M_k$ be $k$ distinct stable marriage solutions. Each man
 $m_i$ has $k$ possible mates under these matchings. Assign him the woman whose rank is
 $j$ among the $k$ (possibly nondistinct) women. Then, this assignment gives rise to another, not necessarily distinct, stable marriage solution.
 \end{theorem}
 
 Their proof uses properties of fractional stable marriages. In this paper, we prove a generalized median theorem on finite distributive lattices.
When this result is applied to the stable matching, we get  Teo and Sethuraman's median stable matching theorem.
Not only our proof is much simpler than \cite{teo1998geometry}, but is also more general because it is applicable to any problem for which
the set of solutions form a finite distributive lattice. For example, the theorem is applicable to  constrained stable matchings \cite{DBLP:journals/corr/abs-1812-10431} such as
stable marriages in which regret of man $m_i$ is at most that of $m_j$, or stable marriages with forbidden pairs.
When the generalized median theorem is applied to the market clearing price vectors \cite{easley2010networks}, we get an analogous result for market clearing prices.

\section{Generalized Median Feasible elements}\label{sec:median}
Let $L$ be any finite distributive lattice. Let $B$ be any boolean predicate defined on $L$. $B$ is called a {\em regular} predicate if
the set of elements of $L$ that satisfy $B$ form a sublattice of $L$  \cite{garg2001slicing}.

In the context of the stable matching problem, we define $L$ as the set of 
all {\em assignment} vectors $G$ of size $n$ such that each component of the vector is a number from $0$ to $n-1$.
The interpretation of $G[i] = j$ is that man $m_i$ is assigned his choice $j$ as his spouse.
The choice $0$ corresponds to the top choice and the choice $n-1$ corresponds to the least preferred choice.
The order defined on $L$ is natural. For any two assignment vectors $G$ and $H$,
 \[ G \leq H \equiv \forall i: G[i] \leq H[i] \]
 In other words, an assignment $G$ is less than or equal to $H$ iff every man gets at least as good a choice
 in $G$ as in $H$.
 
Note that an assignment may not even be a matching if multiple men are assigned the same woman.
We now define the 
predicate $B$ on $L$.
An assignment vector $G$ satisfies $B$ if it is a stable marriage. Formally, $G$ satisfies $B$ if
\begin{enumerate}
\item
 $G$ is a matching, i.e., all men are assigned different woman, and
\item
There is no {\em blocking} pair in $G$. A pair of man and woman $(m,w)$ is blocking if they are not matched and they prefer each other over their
assignment in $G$. 
\end{enumerate}
It is well-known that the set of elements of $L$ that satisfy $B$ form a sublattice of $L$, i.e., if
$G$ and $H$ are stable marriages, then so are $G \sqcup H$ and $G \sqcap H$ where $\sqcup$ and $\sqcap$ are the
join and the meet operation in the lattice $L$. This result is stated by Knuth \cite{knuth1997stable} and attributed to Conway.

\begin{figure}[htbp]
  \centering
   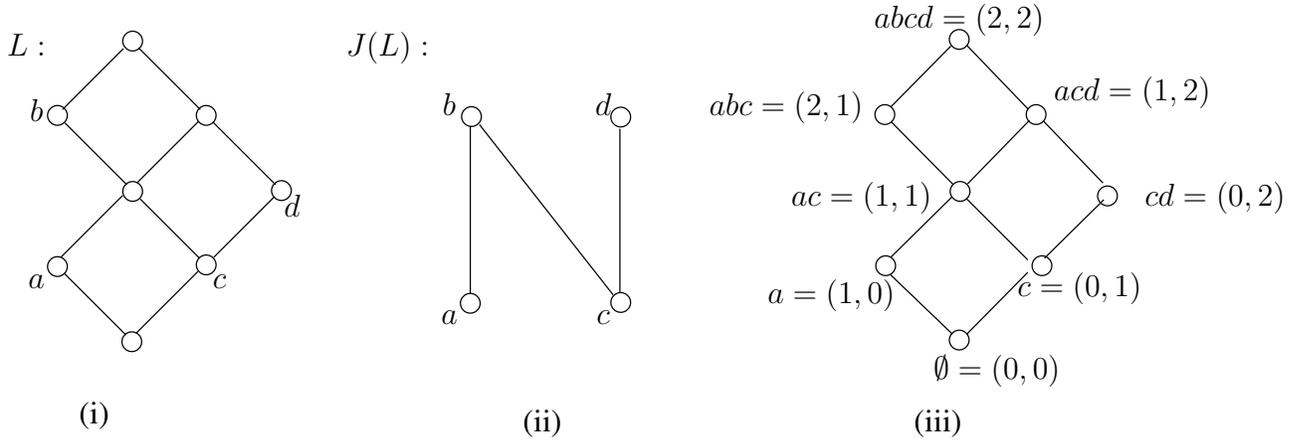
  \caption{(i) A finite distributive lattice $L$ (ii) poset $J(L)$ (iii) Viewing $L$ as order ideals of $J(L)$}
  \label{fig:teo}
\end{figure}

Going back to our general set-up, we have a finite distributive lattice $L$ and a regular predicate $B$ defined on $L$.
Let $M$ be any set of elements in $L$ that satisfy $B$. We now present the general median theorem on finite distributive lattices.
Before we present the theorem, we need to define the notation $G[i]$ for any element $G$ in a general finite distributive lattice $L$.
Let $J(L)$ denote the sub-poset of $L$ consisting of all the join-irreducible elements of $L$. 
Fig. \ref{fig:teo}(i) shows a distributive lattice $L$ and Fig. \ref{fig:teo}(ii) shows the subposet $J(L)$.
Consider any chain partition of $J(L)$. Suppose that $J(L)$ has $n$ chains in its partition.
In our example, $J(L)$ is partitioned into two chains.
Let $H$ be an order ideal of the poset $J(L)$, i.e., $H$ is a subset of $J(L)$ such that
if $y$ belongs to $H$ and $x$ is less than $y$, then $x$ is also included in $H$.
It is clear that $H$ can be equivalently represented as a vector of size $n$ such that $H[i]$
equals the number of elements of chain $i$ of $J(L)$ included in $H$.
In our example, the order ideal $\{a,b,c\}$ is represented as the vector $(2,1)$ because it contains
two elements from the first chain and one element from the second chain.
From Birkhoff's Theorem on finite distributive lattices \cite{Birk3,davey,gar:2015:bk},
there is 1-1 correspondence between order ideals of the poset $J(L)$ and $L$.
Thus, any element $G$ in $L$ is equivalent to an order ideal $H$ of $J(L)$ and
$G[i]$ is same as $H[i]$ which is simply the number of elements of chain $i$ included in $H$.
Fig. \ref{fig:teo}(iii) shows the vector representation of all elements in lattice $L$.
With this notation, we are ready to present the generalized median theorem for finite distributive lattices.

\begin{theorem}\label{thm:median}
Let $L$ be any finite distributive lattice. Let $B$ be any boolean predicate defined on $L$ such that $B$ is regular.
Let $M = \{M_1, M_2, \ldots, M_k\}$ be any subset of $L$ of size $k$ such that all elements of $M$ satisfy $B$. 
Then, for each index $r$, we get a multiset $\{ M_i[r] ~|~1 \leq i \leq k \}$. For any $j$, we construct the general $j$-median state, $G^j$
as follows. For any $r$, $G^j[r]$ is given by the $j^{th}$ element in the {\em sorted} multiset $\{ M_i[r] ~|~1 \leq i \leq k \}$.
Then, $G^j$ also satisfies $B$.
\end{theorem}

Before, we give the proof, we illustrate the theorem on an example.
Consider the lattice $L$ in Fig. \ref{fig:teo}. Suppose that we are given $M = \{(1,0), (0,1), (0,2)\}$ such that all three elements in $M$ satisfy $B$.
When we sort the first index, we get the multiset $\{0,0,1\}$, and when we sort the second index, we get the multiset $\{0,1,2\}$. By using respective components,
the median elements generated from $M$ are $\{(0,0), (0,1), (1,2)\}$. Theorem claims that the median elements also satisfy $B$.
It is easy to verify that the claim is true for this example. The element $(0,0)$ satisfies $B$ because both $(1,0)$ and $(0,1)$ satisfy $B$ which is closed under
meets. The element $(1,2)$ satisfies $B$ because both $(1,0)$ and $(0,2)$ satisfy $B$ which is also closed under joins.

Observe that $k=1$, then Theorem \ref{thm:median} is trivially true for any predicate $B$. When $k=2$, it is true iff $B$ is regular because the generalized
medians correspond to the meet and the join of two elements. Theorem \ref{thm:median} works for any $k$.

We now give the proof.

\begin{proof}
Without loss of generality, we can assume that $M_k$ is a maximal element in the set $M$.
We use induction on $k$. When $k$ equals $1$, the theorem holds trivially because $G_1$ and $M_1$ are identical.
Suppose that theorem holds for all values of $k$ less than $t$. 
Consider the set $M = \{ M_i ~|~ 1 \leq i \leq t-1 \} \cup \{ M_t \}$. 
Let $G_1, G_2, \ldots, G_t$ be median elements of $M$. We show that they satisfy $B$. We have two cases.\\
\\
{\em Case 1}: First suppose that $M_t \geq M_i$ for all $i < t$.\\
In this case, the assertion holds by induction since $G_1, G_2, G_{t-1}$ are the median elements for  $\{ M_i ~|~ 1 \leq i \leq t-1 \}$
and $G_t$ is identical to $M_t$.\\
\\
{\em Case 2}: $M_t$ is not greater than or equal to $M_i$ for some $i < t$.\\
Let $P = \{ P_i  ~|~ 1 \leq i < t \}$ be the median elements for $\{ M_i ~|~ 1 \leq i < t \}$.
From the induction hypothesis, we can assume that all elements in $P_i$ satisfy $B$.
From the definition of the median elements, we know that $P_{t-1} \geq P_i$ for all $1 \leq i \leq t-1$.
We construct $P'$, a set of size $t$, from $P$ as follows:
\[ P' := \{ P_1, P_2, \ldots, P_{t-2}, P_{t-1} \sqcap M_t, P_{t-1} \sqcup M_t \}. \]
Since $t>1$, we are guaranteed that $P_{t-1}$ exists and therefore the construction is valid.

We first claim that the set of median elements for $M$ and $P'$ are identical.
Let $Q$ be defined as $P \cup \{ M_t \}$. It is clear that the median elements of $M$ and $Q$ are identical
because $P$ is the set of median elements of  $\{ M_i ~|~ 1 \leq i \leq t-1 \}$. We 
show that the median elements of $P'$ and $Q$ are identical.
Each of $P_i$ for $1 \leq i \leq t-2$ contributes exactly once in the multiset $P'[j]$ and the multiset $Q[j]$ for any $j$.
Since the last two elements in $P'$ are $P_{t-1} \sqcap M_t, P_{t-1} \sqcup M_t$, we get that 
$P_{t-1}$ and $M_t$ also contribute exactly once for $P'[j]$ and $Q[j]$.
Hence, median elements of $P'$ and $Q$ are identical which implies that the median elements of $P'$ and
$M$ are identical.

We next claim that the median elements of $P'$ satisfy $B$.
Since, $P_{t-1} \sqcup M_t$ is greater than or equal to all elements in $P'$, case 1 is applicable and 
we get that all the median elements of $P'$ satisfy $B$.
\end{proof}

By using $L$ as the set of all assignments to men and $B$ as the predicate that the assignment is a stable marriage, we get the following
Teo and Sethurmanan's result on generalized median stable matching  \cite{teo1998geometry}.
\begin{corollary}
Let $M = \{M_1, M_2, \ldots, M_k\}$ be any set of stable marriages for any instance of the stable marriage problem.
For any $j$, if every man is assigned the $j^{th}$ element in the {\em sorted} multiset $\{ M_i[m] ~|~1 \leq i \leq k \}$, then
the resulting assignment is a stable marriage.
\end{corollary} 

Moreover, we get the following result on market clearing prices.
\begin{corollary}
 Let $M = \{M_1, M_2, \ldots, M_k\}$ be any set of market clearing prices for any instance of the market clearing prices problem.
For any $j$, if every item is assigned the $j^{th}$ price in the {\em sorted} multiset $\{ M_i[m] ~|~1 \leq i \leq k \}$, then
the resulting price assignment is market clearing.
\end{corollary}

\section{Acknowledgements}
I thank Changyong Hu and Xiong Zheng for some discussions on the topic.
\bibliography{refs,refs2,refs3,myrefs}

\end{document}